\documentclass[sigconf]{acmart}

\setcopyright{none}
\renewcommand\footnotetextcopyrightpermission[1]{} 
\AtBeginDocument{%
  }

\copyrightyear{2018}
\acmYear{2018}
\acmDOI{XXXXXXX.XXXXXXX}
\acmConference[Conference acronym 'XX]{Make sure to enter the correct
  conference title from your rights confirmation emai}{June 03--05,
  2018}{Woodstock, NY}
\acmISBN{978-1-4503-XXXX-X/18/06}

\usepackage{xspace}
\usepackage{subfigure}
\usepackage{caption}
\usepackage{graphicx}
\usepackage{float} 
\usepackage{subcaption}

\usepackage{circledsteps}

\newcommand{\hi}[1]{\vspace{.25em} \noindent {\bf #1}\xspace}

\newcommand{\oursys}{\texttt{FeatInsight}\xspace}

\newcommand{\blue}[1]{\textcolor{blue}{#1}}

\setcopyright{none}
\renewcommand\footnotetextcopyrightpermission[1]{} 
\begin{document}

\title{FeatInsight: An Online ML Feature Management System\\ on 4Paradigm Sage-Studio Platform}

\author{Xin Tong}
\affiliation{\institution{{Shanghai Jiao Tong Univ.}}\country{}}
\email{{culily02@gmail.com}}

\author{Xuanhe Zhou}
\affiliation{\institution{{Shanghai Jiao Tong Univ.}}\country{}}
\email{{zhouxh@cs.sjtu.edu.cn}}

\author{Bingsheng He}
\affiliation{\institution{{National Univ. of Singapore}}\country{}}
\email{{hebs@comp.nus.edu.sg}}

\author{Guoliang Li}
\affiliation{\institution{{Tsinghua University}}\country{}}
\email{{liguoliang@tsinghua.edu.cn}}

\author{Zirui Tang}
\affiliation{\institution{{Shanghai Jiao Tong Univ.}}\country{}}
\email{{raymondtangzirui@163.com}}

\author{Wei Zhou}
\affiliation{\institution{{Shanghai Jiao Tong Univ.}}\country{}}
\email{{weizhoudb@gmail.com}}

\author{Fan Wu}
\affiliation{\institution{{Shanghai Jiao Tong Univ.}}\country{}}
\email{{fwu@cs.sjtu.edu.cn}}

\author{Mian Lu}
\affiliation{\institution{{4Paradigm Inc.}}\country{}}
\email{{lumian@4paradigm.com}}

\author{Yuqiang Chen}
\affiliation{\institution{{4Paradigm Inc.}}\country{}}
\email{{chenyuqiang@4paradigm.com}}

\renewcommand{\shortauthors}{Xin Tong et al.}

\begin{abstract}
Feature management is essential for many online machine learning applications and can often become the performance bottleneck (e.g., taking up to 70\% of the overall latency in sales prediction service). Improper feature configurations (e.g., introducing too many irrelevant features) can severely undermine the model’s generalization capabilities. However, managing online ML features is challenging due to (1) large-scale, complex raw data (e.g., the 2018 PHM dataset contains 17 tables and dozens to hundreds of columns), (2) the need for high-performance, consistent computation of interdependent features with complex patterns, and (3) the requirement for rapid updates and deployments to accommodate real-time data changes. In this demo, we present \oursys, a system that supports the entire feature lifecycle, including feature design, storage, visualization, computation, verification, and lineage management. \oursys (with OpenMLDB as the execution engine) has been deployed in over 100 real-world scenarios on 4Paradigm’s Sage Studio platform, handling up to a trillion-dimensional feature space and enabling millisecond-level feature updates. We demonstrate how \oursys enhances feature design efficiency (e.g., for online product recommendation) and improve feature computation performance (e.g., for online fraud detection). The code is available at \emph{\blue{\url{https://github.com/4paradigm/FeatInsight}}}.
\end{abstract}

\ccsdesc[500]{Do Not Use This Code~Generate the Correct Terms for Your Paper}
\ccsdesc[300]{Do Not Use This Code~Generate the Correct Terms for Your Paper}
\ccsdesc{Do Not Use This Code~Generate the Correct Terms for Your Paper}
\ccsdesc[100]{Do Not Use This Code~Generate the Correct Terms for Your Paper}



\maketitle

\vspace{-.5em}

\section{INTRODUCTION}

\begin{figure}[!t]
  \vspace{.5em}
  \centering
  \includegraphics[width=\linewidth]{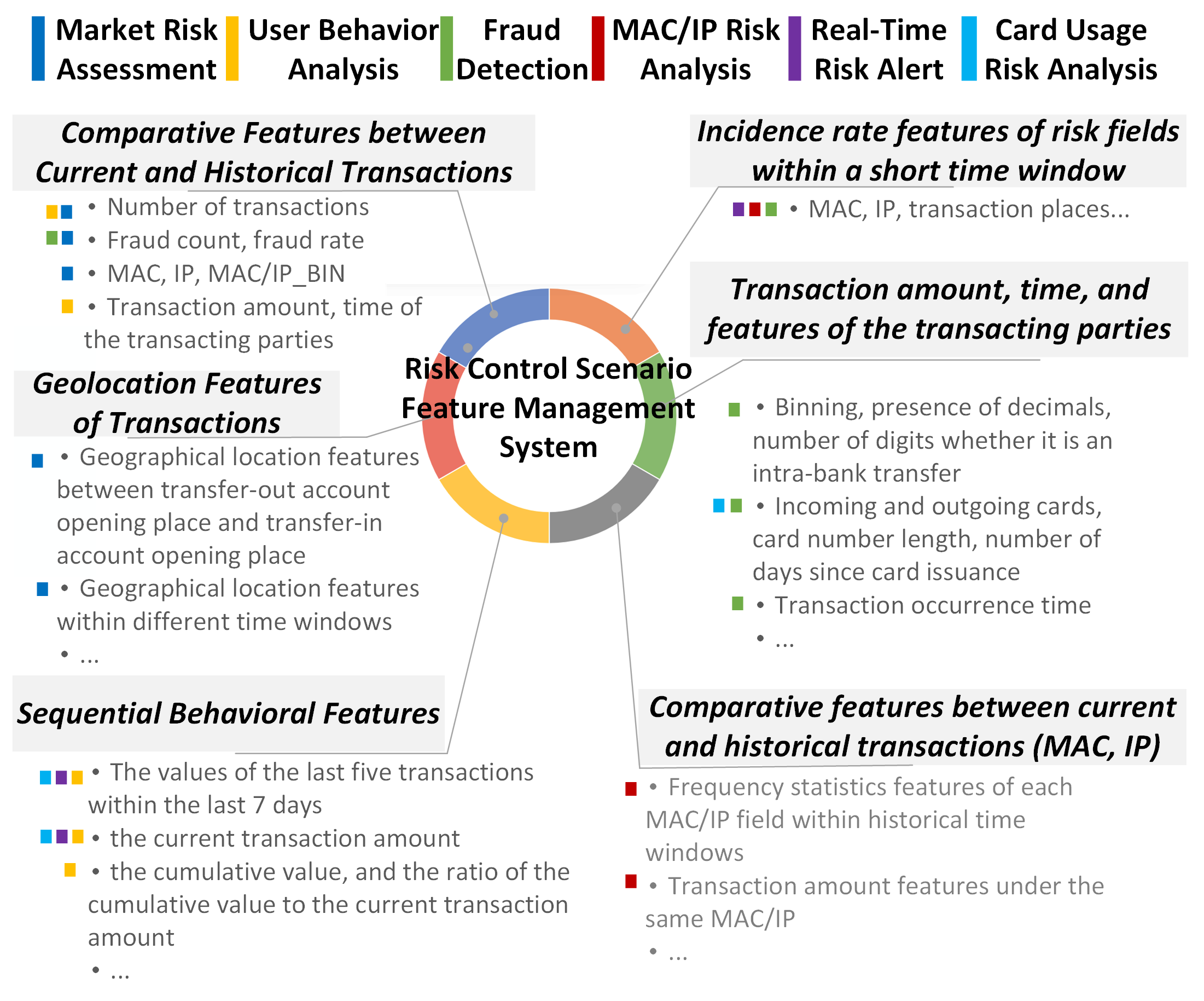}
  \vspace{-2.5em}
  \caption{Online ML Features \textnormal{-- The complexity of managing online machine learning features that span over six categories and involve 600+ computation patterns in a risk control scenario~\cite{TheCreditScoringToolkit}.}}
  \vspace{-2em}  
  \label{riskScenario}
\end{figure}

Feature management lies at the heart of modern machine learning (ML) development and deployment, covering tasks ranging from data ingestion, analysis, storage, and transformation to model training, optimization, and final deployment~\cite{CAI201870}. For instance, in risk control scenarios (see Figure~\ref{riskScenario}), ML-based functions such as market risk assessment and card usage risk analysis often rely on \emph{more than 600 features} covering multiple categories (e.g., geo-locations, incidence rates, and time-series features). These features can require specialized computation patterns (e.g., long-window computations, multi-window joins)~\cite{febench}.

Although there are prior works on feature selection~\cite{10.1007/978-3-662-44320-0_12,autocross}, feature computation~\cite{Zhang2023ScalableOI,mezati2024flink}, and feature augmentation~\cite{FeatureAugmentation1}, an end-to-end feature management system remains lacking. These works excel at either offline feature generation or online inference but seldom unify both, complicating automation of feature pipelines, lineage management, reliable reuse, and consistency checks across different ML workflows. Instead, an effective feature management system should address the following considerations. \emph{(1) Visual and SQL-Assisted Feature Design:} The system should provide an intuitive, user-friendly interface covering the entire feature lifecycle. \emph{(2) Consistent Feature Computation:} It must ensure consistent computation across offline and online pipelines, minimizing deployment overhead. \emph{(3) High-Performance Feature Management:} It should offer efficient storage and optimized data-handling strategies to support large-scale feature computation with minimal latency.

However, designing such a system poses several key challenges. \emph{(1) High Dimensionality:} Hundreds or even thousands of features may be sourced from multiple data tables, causing large feature spaces. Raw datasets can also be highly interrelated (e.g., three tables with 828 columns in the IEEE-CIS Fraud Detection scenario yield 393 features, some involving multi-column calculations~\cite{ieee_cis_fraud_detection}). \emph{(2) Complex Interdependencies:} Data contributing to final features often exhibits complex relationships, making it difficult to analyze without user-friendly tools. Moreover, aligning features across different execution engines can necessitate costly consistency checks before deployment, particularly when features rely on complex or interdependent data. \emph{(3) Dynamic Changes:} Frequent updates and redeployments are required to accommodate shifting scenarios (e.g., version upgrades), demanding rapid deployment without compromising performance or consistency.

\begin{sloppypar}
To address these challenges, we introduce \oursys, a comprehensive feature management system offering three main capabilities. First, \oursys provides a user-friendly visual interface for feature design, storage, computation, and lineage management. Users can easily build ML features via a visual, drag-and-drop SQL tool and then deploy them with a single click, automating everything from SQL creation to large-scale deployment. Similarly, platforms like \emph{4Paradigm Sage Studio} offer a visual, drag-and-drop interface for AI application development, simplifying the process of building and deploying machine learning models.  Second, \oursys supports efficient, versatile feature computation with built-in consistency verification. It leverages high-performance execution engines (OpenMLDB~\cite{openmldb}) for both online and offline feature computations and provides an interface to ensure consistent results. Specialized ML functions (e.g., top-$N$ frequency counts) and feature signatures for high-dimensional scenarios (e.g., labeling product-item features) further enhance flexibility. Third, \oursys employs a compact in-memory data storage format and stream-optimized data structures, pre-sorting data by key and timestamp for rapid online access under constrained resources.
\end{sloppypar}

In this demonstration, we introduce the workflow of \oursys, showing how to design, store, visualize, and export features for offline training, and then deploy them as an online feature service. We also highlight the practice of \oursys in scenarios like online product recommendation and online fraud detection. Beyond these, \oursys has been successfully applied in domains like social media, healthcare, and energy prediction~\cite{febench,openmldb}.



\vspace{-.75em}
\section{\texttt{FEATINSIGHT} OVERVIEW}
\vspace{-.25em}

\begin{figure}[!t]
  \centering
  \includegraphics[width=.98\linewidth]{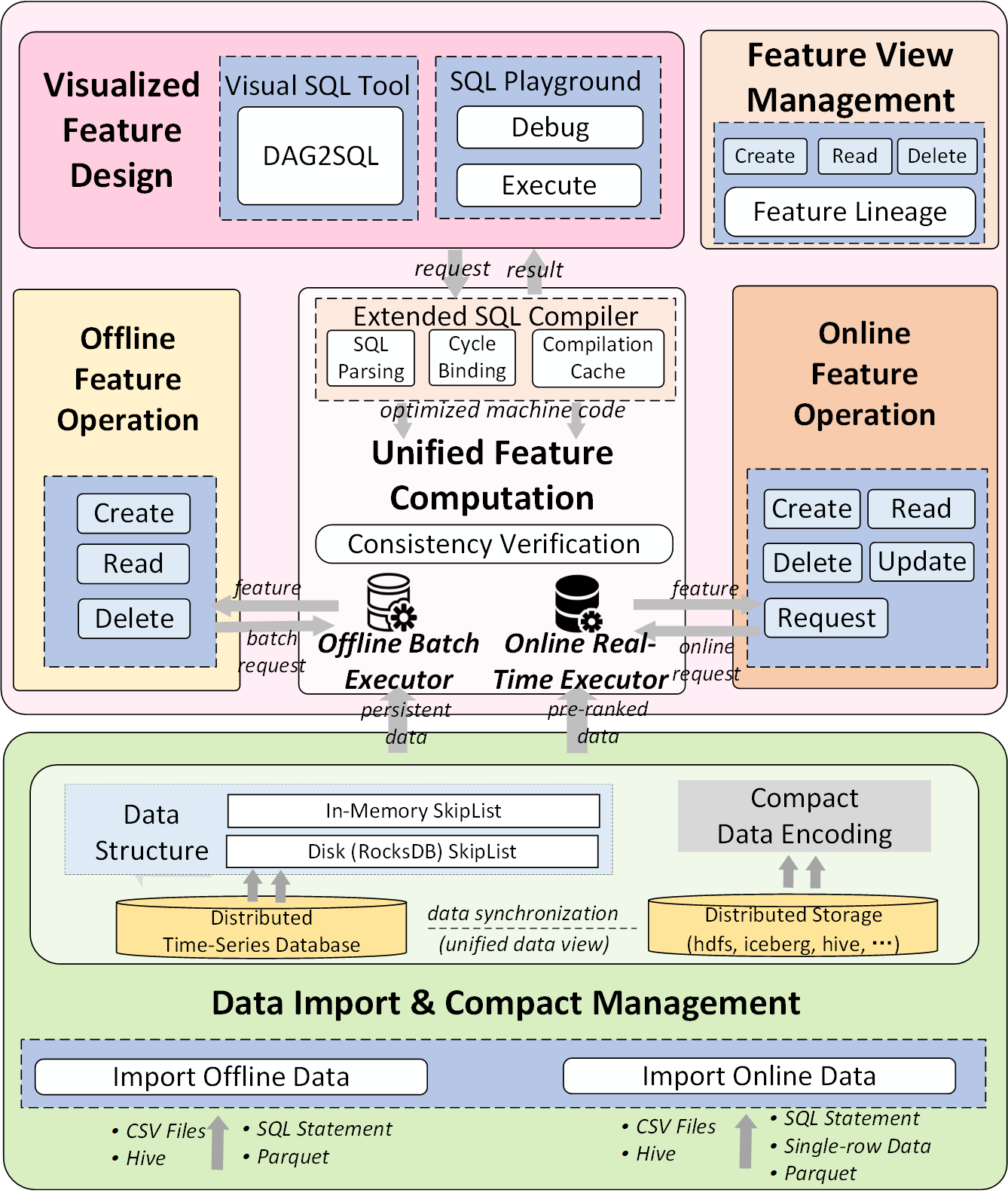}
  \vspace{-1.25em}
  \caption{The Architecture of \oursys.}
  \label{fig:arch}
  \vspace{-3.5em}
\end{figure}

As shown in Figure~\ref{fig:arch}, \oursys comprises four main components, including visualized feature design, feature view management, unified feature computation, and data import \& compact time-series data management.

\hi{Visualized Feature Design.} To simplify the process of constructing SQL statements, we first provide a visual feature design tool. Users can connect graphical SQL blocks to represent hierarchical relationships, and \oursys transforms the resulting Directed Acyclic Graph (DAG) into executable SQL statements. Leveraging OpenMLDB’s SQL execution engines~\cite{openmldb}, \oursys supports both debugging and execution of the generated SQL statements (SQL playground), allowing users to perform arbitrary SQL operations and debug feature computation routines. For well-verified SQL scripts, \oursys compiles them into efficient C++ machine code by applying optimizations such as parsing optimization, cyclic binding, and compilation caching.

\hi{Feature View Management.} In \oursys, we utilize \emph{feature views} to collectively group and manage features with similar usage patterns, where each feature view corresponds to a set of features defined by a single SQL computation statement. Based on the definition of feature views, we support feature lineage management. That is, we link each feature to its corresponding feature view, database, and construction SQL, making it easier to trace the evolution of feature definitions. Additionally, in the online feature service, \oursys uses caching strategy to store earlier versions of feature services, enabling users to reuse prior SQL scripts and incrementally add new raw data attributes.

\hi{Unified Feature Computation.} \oursys employs execution engines optimized for ML scenarios. (1) For offline data, it enhances resource utilization by parallelizing window operations on the same tables and mitigates data skew by dynamically reassigning window data according to key columns and data distribution. (2) For online data, we apply \textit{pre-aggregation} to handle long time intervals (e.g., for years) or hotspot data and \textit{dynamic data adjusting} to efficiently handle out-of-order streams over multiple tables~\cite{Zhang2023ScalableOI}. To ensure high availability and fault tolerance, \oursys integrates Zookeeper (a distributed framework~\cite{267345}). (3) We support feature consistency verification. That is, we perform feature computation of test data through execution engines in both offline and online scenario, and compare the consistency of the result. 

\hi{Data Import \& Compact Management.} \oursys imports raw data in multiple formats, including CSV, Hive, SQL statements, Parquet, and single-row data. To reduce storage overhead and manage large datasets under limited memory resources, \oursys employs a compact encoding format that reduces memory usage by combining fixed-size and variable-size data structures. Additionally, it mitigates lock contention and rebalancing overhead through lock-free reads and writes with atomic compare-and-swap (CAS) operations. Outdated data is efficiently removed via timestamp ordering and batch deletion, helping maintain optimal performance and real-time freshness.



\begin{figure*}[htbp] 
    \vspace{-.25em}
    \centering
    \includegraphics[width=.98\textwidth]{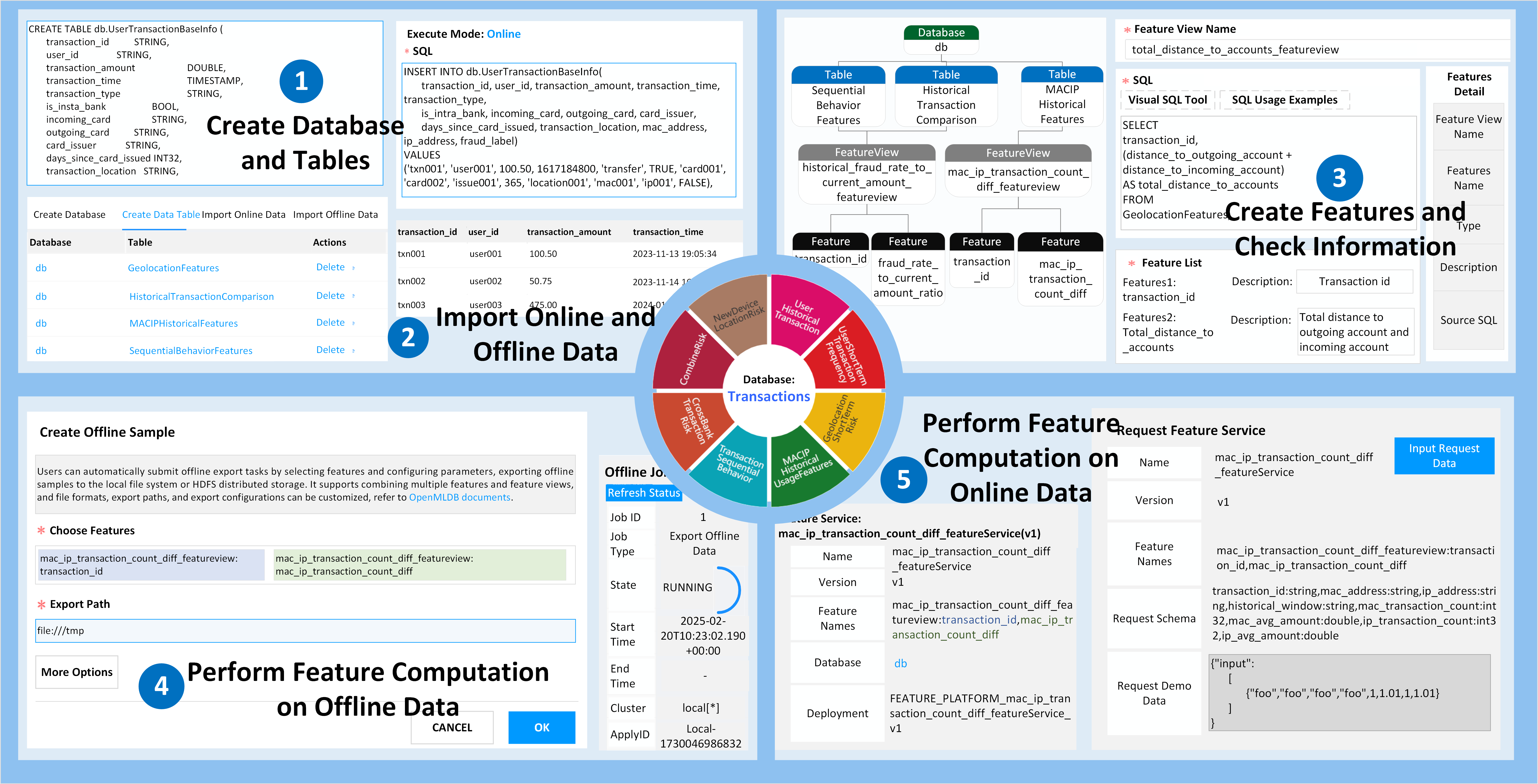} 
    \vspace{-.25em}
    \caption{A Screenshot and Running Example of \oursys.}
    \label{fig:scenario}
\end{figure*}

\begin{figure}[h]
  \vspace{-.75em}
  \centering
  \includegraphics[width=1\linewidth]{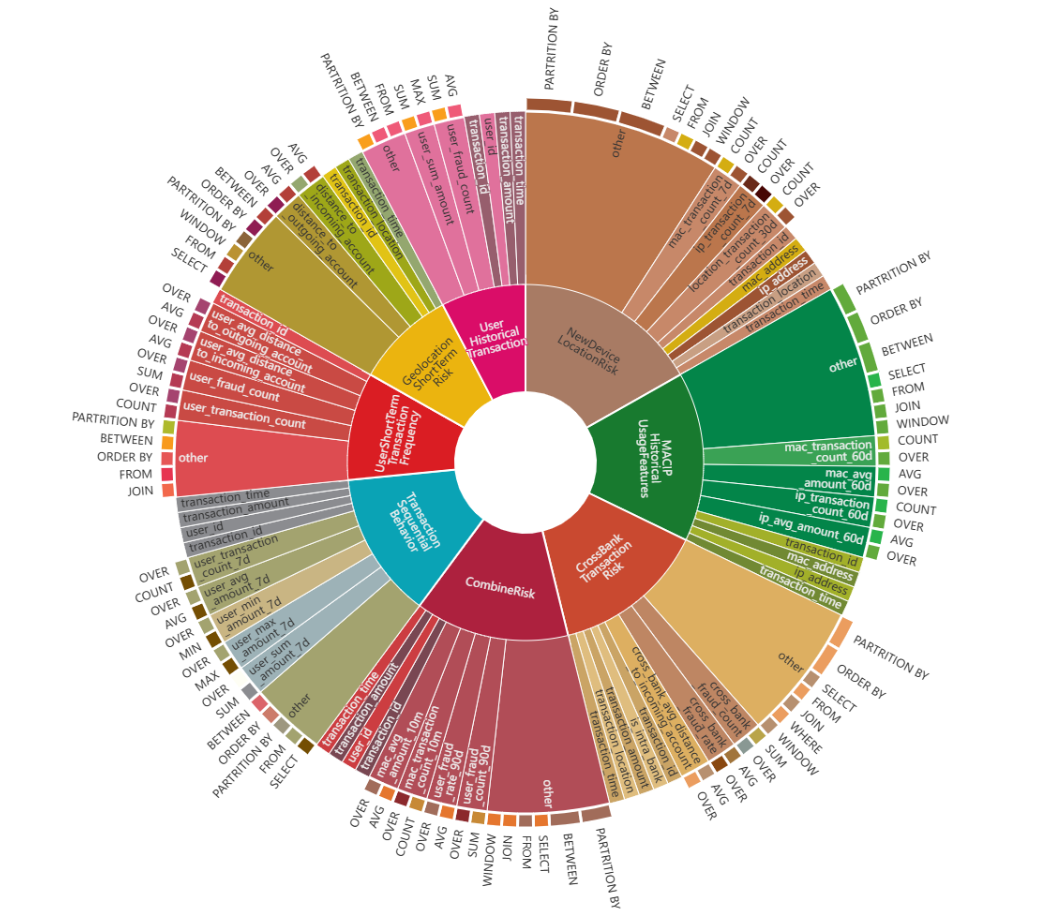}
  \vspace{-1.5em}
  \caption{The Distribution of 784 Features (designed within \oursys) for Fraud Detection in Banking Scenario.}
  \label{manyFeatures}
  \vspace{-2.5em}
\end{figure}

\section{DEMONSTRATION}

\begin{sloppypar}
In this section, we first demonstrate the user interface of \oursys, illustrating how users can easily import offline and online data in various formats, create and update features, and efficiently compute these features in both offline and online modes. Subsequently, we present the practical outcomes of deploying \oursys in real-world online machine learning applications. 
\end{sloppypar}

\vspace{1.5em}
\subsection{End-to-End Experience.}



Figure~\ref{fig:scenario} is a screenshot of the front-end of \oursys. The user can manage raw data (in relational tables) and features in both offline and online scenarios with the following steps.

\noindent \underline{{\it 1) Import Data}}. Before creating features, users first need to create a database, define data tables, and import both online and offline data. To begin, users click the ``Data Import'' button to create a new database. Next, they can choose to create a data table via the visualization interface or through other methods such as SQL, Parquet, and Hive (Figure~\ref{fig:scenario}-\Circled{1}). Once the data table is established, users can load data for both online and offline scenarios. With declarative tools, they can use the SQL commands \verb|LOAD DATA| and \verb|INSERT|, or import data from preexisting CSV files. Finally, users wait for the task to complete, monitoring its status and logs in real time (Figure~\ref{fig:scenario}-\Circled{2}).

\begin{sloppypar}
\noindent \underline{{{\it 2) Create Features}}}. Next users can create features (Figure~\ref{fig:scenario}-\Circled{3}). The user should fill in the form accordingly, including the feature view name, database name, SQL to create basic features. Description of the features can be filled in at the same time. After clicking the "Submit" button, the feature is saved and then users can check its details. By clicking on the name, users can view its basic information including feature view name, feature name, type, description and source SQL. Besides, the feature values can also be previewed.
\end{sloppypar}

\begin{sloppypar}
\noindent \underline{{{\it 3) Perform Feature Computation on Offline Data}}}. After the {\it Features Creating} procedure, the user can perform feature computation on offline data and export training dataset for machine learning models (Figure~\ref{fig:scenario}-\Circled{4}). Firstly, the user can choose the features to export and specify the export path. There are ``More Options'' to specify the file format and other advanced parameters. After that, users can check the status at ``Offline Samples''. Finally, the detailed exported samples can be checked from the data files in the export path. 
\end{sloppypar}

\begin{sloppypar}
\noindent \underline{{{\it 4) Perform Feature Computation on Online Data}}}. Besides checking the offline samples, users can use the Online Feature Service in \oursys (Figure~\ref{fig:scenario}-\Circled{5}). Users can name a new feature service, fill in its version, choose the features to deploy and add feature service description. After successful creation, users can check service details, including its SQL, features list, dependent tables and lineage. Additionally, we also provide the ``Request Feature Service'' page, where users can choose specific feature service to test its functionality and performance by computing over test data.
\end{sloppypar}

\vspace{-.75em}
\subsection{Scenario 1 - Efficient Feature Deployment for Online Product Recommendation}
\begin{figure}[h]
  \vspace{-1.25em}
  \centering
  \includegraphics[width=.98\linewidth]{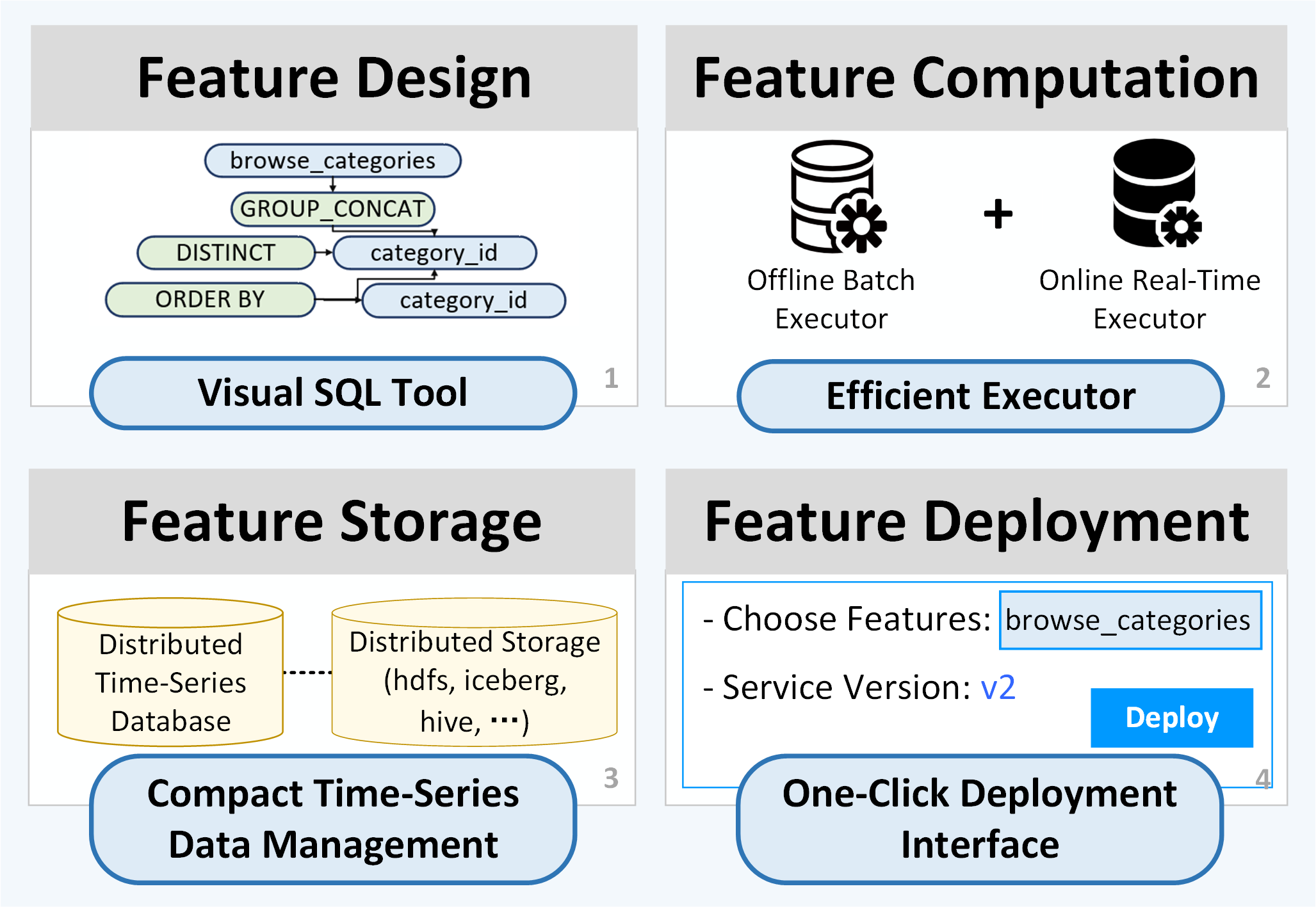}
  \vspace{-.5em}
  \caption{Efficient Feature Deployment by \oursys.}
  \label{feature-deployment}
  \vspace{-1.5em}
\end{figure}

\begin{sloppypar}
\oursys~{can significantly enhance the feature deployment efficiency}. Taking online product recommendation in \emph{Vipshop}  as an example. In this scenario, data is updated on a minute-by-minute basis, with a total of 720 million order records updated daily. To achieve real-time personalized product recommendations, it is crucial to deploy features quickly.

\oursys achieve efficient feature deployment through four optimization: (1) \oursys supports visualized feature design such as SQL drag-and-drop tool, making it easier for non-professionals to design features. (2) \oursys utilizes unified executors, improving the efficiency of feature computation. Besides, we supports consistency verification between offline and online samples, reducing the time spent on costly verification, which based on our experience (e.g., at Akulaku \cite{akulaku_openmldb}) can take several months or even one year. (3) the compact time-series data management ensures rapid online data access. (4) \oursys packages all the steps for feature deployment, allowing users to simply click a button, with all steps executed in the background. Through these optimization, \oursys ensures that features can be deployed within an hour, reducing the entire feature development-to-deployment process to just five person-days (improving the efficiency by over 60\%). 


\end{sloppypar}

\subsection{Scenario 2 - Rapid Feature Computation\\ for Online Fraud Detection}

\begin{sloppypar}
Moreover, \oursys supports millisecond-level feature computation. For the fraud detection task at a \emph{major Fortune Global 500 bank}, the historical dataset comprises roughly 100 million customer records, 500 million card records, and 1.7 billion transactions, with a daily volume of about 4.5 million transactions. Online fraud detection requires identifying potential fraud through the computation of critical features (e.g., transaction amounts, frequencies, geo-locations, user behaviors, and device information). We observe that (1) a naive Spark system achieves around 200\,ms response time but suffers from relatively poor recall; (2) a customized Spark version developed in-house delivers roughly 50\,ms response time yet provides only moderate recall; and (3) our \oursys-based solution achieves sub-20\,ms response (with QPS exceeding 1000) while maintaining higher recall.

On the one hand, as shown in Figure~\ref{manyFeatures}, users design over 784 online features with \oursys, including seven-day transaction aggregations and MAC/IP address details, which directly boost fraud detection recall. On the other hand, the key to high performance lies in a refined skiplist data structure sorted by both key and timestamp, combined with a lightweight row-encoding scheme that enables lock-free read/write operations—allowing on-the-fly feature updates without costly contention. By leveraging OpenMLDB’s execution engine, \oursys efficiently transforms incoming transaction data into real-time features, maintaining partial aggregators (e.g., rolling sums and distinct counts) in memory so new data can be quickly merged with the existing state. Specialized ML functions, such as top-$N$ frequency counts and feature signatures, handle high-dimensional data (e.g., labeling sparse or dense feature columns) and ensure consistent results across both online and offline pipelines. Collectively, these optimizations enable \oursys to accommodate rapid data updates while delivering real-time fraud detection under strict latency constraints.

\end{sloppypar}

\bibliographystyle{ACM-Reference-Format}
\bibliography{sample-base}


\end{document}